# Criticality in the brain: A synthesis of neurobiology, models and cognition

Luca Cocchi[1*], Leonardo L. Gollo[1], Andrew Zalesky[2], Michael Breakspear[1]

[1] Queensland Institute for Medical Research, Brisbane, Australia.
[2] Melbourne Neuropsychiatry Centre, The University of Melbourne, Melbourne, Australia.

Abstract

Cognitive function requires the coordination of neural activity across many scales, from neurons and circuits to large-scale networks. As such, it is unlikely that an explanatory framework focused upon any single scale will yield a comprehensive theory of brain activity and cognitive function. Modelling and analysis methods for neuroscience should aim to accommodate multiscale phenomena. Emerging research now suggests that multi-scale processes in the brain arise from so-called critical phenomena that occur very broadly in the natural world. Criticality arises in complex systems perched between order and disorder, and is marked by fluctuations that do not have any privileged spatial or temporal scale. We review the core nature of criticality, the evidence supporting its role in neural systems and its explanatory potential in brain health and disease.

Highlights

- Criticality is a wide-spread phenomenon in natural systems
- Criticality provides a unifying framework to model and understand brain activity and cognitive function
- Substantial evidence now supports the hypothesis that the brain operates near criticality
- We review the role of criticality in healthy and pathological brain dynamics
- Caveats and pitfalls regarding the assessment of criticality in the brain are discussed

* Email: Luca.Cocchi@qimrberghofer.edu.au

# List of abbreviations

**ATP:** Adenosine Triphosphate

**BS**: Burst suppression

**DC:** Direct current

**DFA:** Detrended fluctuation analysis

**ECoG:** Electrocorticography

**EEG:** Electroencephalogram

**fMRI:** Functional magnetic resonance imaging

**GABA:** Gamma-Aminobutyric acid

**IBI:** Inter-burst interval

**MEG:** Magnetoencephalography

**REM:** Rapid eye movement sleep

**SOC:** Self-organised criticality

**SWS:** Slow-wave sleep

**T$_c$:** Critical temperature

# Table of Contents

# 1. Introduction

Enormous strides have been achieved in neuroscience across a hierarchy of scales of enquiry, from the variety of neural cell types and their molecular biology, through the function of cortical circuits and, in recent years, to the complex architecture of large-scale brain networks. Much of this success has been achieved within research silos, with a focus on scale-specific phenomena, partly mandated by the apertures of various imaging technologies and partly by the training and cultures within the various neuroscientific disciplines. Research in neuroscience also proceeds within a largely descriptive world-view, with increasing emphasis on the collation and statistical characterization of "big data" (Biswal *et al.*, 2010; Markram *et al.*, 2015). Whilst specific mechanisms have been elucidated across an array of basic and clinical neuroscience domains, important challenges remain to be addressed: First, since correlations between behaviour and neuronal activity have been documented at almost every scale of analysis, it seems unlikely that a description of the brain at any particular scale will be sufficient to describe brain function. How is neural activity integrated across scales to give rise to cognitive function? What are the mechanisms linking activity across scales? Second, brain function does not only rely upon the execution of particular functions, but also on adaptive switching from one function to another, depending on context and goals. What are the fundamental principles underlying such complex, flexible neuronal dynamics? Third, what are the major theoretical frameworks to explain and unify the properties of all the large volumes of data currently being accrued? Fourth, how is information encoded by neurons – in the entropy of individual spikes, or via likelihood functions encoded by the distributions of population activity?

The principles that unify brain function across spatial and temporal scales remain largely unknown. However, comparable multi-scale challenges exist in other scientific disciplines. Meteorology, for example, spans scales from local wind gusts through regional weather systems up to global climate patterns. Each scale is nested within a larger scale, such that the local variance in wind gusts depends upon the regional weather, which is likewise constrained by global trends such as *El Niño*. Mathematicians and physicists have developed a considerable armoury of analytic tools to address multi-scale dynamics in a host of physical, biological and chemical systems (Bak *et al.*, 1987). Chief amongst these is the notion of *criticality,* an umbrella term that denotes the behaviour of a system perched between order (such as slow, laminar fluid flow) and disorder [such as the turbulence of a fast-flowing fluid, (Shih *et al.*, 2015)]. A critical system shows *scale-free* fluctuations that stretch from the smallest to the largest scale, and which may spontaneously jump between different spatiotemporal patterns. Despite their apparent random nature, the fluctuations in these systems are highly structured, obeying deep physical principles that show commonality from one system to the other (so-called *universality*). They can hence be subject to robust statistical analysis and modelling.

Critical systems thus display the type of cross-scale effects and dynamic instabilities linking activity at different scales that is typical of brain functioning. An emerging literature suggests that brain function may be supported by critical neural dynamics, with original research that continues to flourish (Deco and Jirsa, 2012; Kelso *et al.*, 1992; Priesemann *et al.*, 2014; Scott *et al.*, 2014) on the background of an existing body of reviews and syntheses (Beggs and Timme, 2012; Chialvo, 2010; Deco and Jirsa, 2012; He, 2013; Hesse and Gross, 2015; Kelso *et al.*, 1992; Plenz and Thiagarajan, 2007; Priesemann *et al.*, 2014; Schuster *et al.*, 2014; Scott *et al.*, 2014; Shew and Plenz, 2013). The principles supporting the emergence of these patterns of activity are not yet fully understood but recent studies using neuroimaging techniques such as functional magnetic resonance imaging (fMRI) and electroencephalogram (EEG) (Deco *et al.*, 2009; Linkenkaer-Hansen *et al.*, 2001; Stam and de Bruin, 2004) have added to earlier work in slice preparations (Beggs and Plenz 2003). Computational models also show that neural systems have maximum adaptability to accommodate incoming processing demands when they are close to a critical point (Friston *et al.*, 2012b; Friston, 2000; Gollo and Breakspear, 2014; Kastner *et al.*, 2015; Shew *et al.*, 2009; Yang *et al.*, 2012). Conversely, brain disorders, as diverse as epilepsy, encephalopathy, bipolar disorder and schizophrenia may correspond to excursions from such an optimal critical point.

Despite the ubiquity of criticality in many branches of science, its application to neuroscience is relatively recent and unknown to many neuroscientists. When it is used, it is often invoked metaphorically; a practice which risks mixing distinct processes incorrectly into a rubric term. Research into criticality has much to offer neuroscientists, but needs to be used in accordance with its well-defined operational criteria. Accumulating evidence should also be viewed cautiously according to emerging pitfalls. Here, we first revisit the core notion of critical phenomenon and provide examples from the physical sciences. We then review the classic and recent studies of neuronal criticality. We finally consider emerging applications that advance new theories of healthy and maladaptive cognition using the innovative tools that criticality provides.

## 2. Criticality in physical systems

*Criticality* refers to the appearance of erratic fluctuations in a dynamical system that is close to losing dynamic stability. Because the nature of the instability can vary (as we review below), criticality is a broad umbrella term that subsumes several related phenomena but also excludes others. In this section, we present a brief pedagogical account of criticality. We first consider critical fluctuations that occur close to instability in systems consisting of only a few interacting components. This allows us to introduce core signatures of criticality; namely the emergence of *scale-free* temporal fluctuations, *slowing down* and *multistability,* defined below. We then consider criticality in complex systems composed of many interacting parts. These extra degrees of freedom allow for the occurrence of scale-free spatiotemporal fluctuations known as *avalanches.* Finally, we consider *self-organized criticality* - that is, the process by which criticality emerges without the need for external tuning of a control parameter.

### 2.1. Criticality and bifurcations

We first consider dynamical systems composed of only a few interacting components. Consider the classic example in which two species interact as predator and prey. When the interactions amongst the species are weak and alternative food sources are available, relatively simple models predict that the populations of both species reach stable equilibria, and the processes of consumption and reproduction occur at a steady rate (Berryman, 1992). However, if the interactions between the species increase (i.e. the predators rely more heavily upon the prey population), there reaches a *critical point of interactivity* above which the two populations begin to oscillate: When the predator population is relatively low, the number of prey animals grows through unbalanced reproduction. This then yields a ready food source for the surviving predators, whose population increases. However, as the increasing numbers of predators consume the available prey species, the numbers of the latter then decline, with a subsequent effect on the survival of the predators; the cycle then begins anew. This transition from steady state to cyclic behaviour due to strong interactions is called a *bifurcation* (**Figure 1a**). The strength of the interaction is called a *control parameter* and the point at which the bifurcation occurs is denoted the *critical point*. For three or more interacting species, further bifurcations to more complex dynamics can occur, leading from periodic to chaotic oscillations (smooth and deterministic but aperiodic oscillations) (Arneodo *et al.*, 1980; Vano *et al.*, 2006).

There are two crucial variations on this simple example. First physical processes, such as the predator-prey example, inevitably occur in the presence of small but unceasing random fluctuations. This noise arises from a myriad of causes such as the probabilistic nature of individual predator-prey encounters as well as influences not explicitly modelled (diseases, fluctuating environmental conditions etc). When the interactions between predator and prey are weak, the equilibrium state is very stable and the presence of such random fluctuations only have a minor impact on the observed steady state populations. Likewise, if the interactions are strong, the cyclic oscillations in population numbers are also very stable: The amplitude of the oscillations is relatively stable and the noise is again effectively suppressed. More technically, away from the critical point, the system is said to be *dynamically stable* and the fluctuations are strongly damped, dropping off quickly (with an exponential decay rate). However, in the immediate vicinity of the critical point, the perturbations grow in magnitude, dominating the observations because the system is less stable (or *weakly stable)*. That is, the variance of the observed fluctuations grows in magnitude. Moreover, the fluctuations decay slowly. To be precise, the decay of fluctuations in time changes from a fast (exponential, **Figure 1b**, inset) to a slow (power-law, **Figure 1c**, inset) process. Fluctuations with power-law correlations are *scale-free* because they do not have a characteristic time scale. Upon further increases in the control parameter, the fluctuations in the envelope of the oscillations quickly become stable (**Figure 1d**) and drop off quickly (exponentially, likewise the inset of **Figure 1b**).

Hence near the critical point of a bifurcation, we encounter two central features of criticality: High amplitude scale-free fluctuations (**Figure 1h**) and *slowing down* (i.e., longer autocorrelation, **Figure 1i**). These large, slow fluctuations are termed *crackling noise* after the sound they make if played audibly (Sethna *et al.*, 2001).

A second variation concerns the nature of the bifurcation itself. The predator-prey model exhibits a classic bifurcation, whereby a single critical point separates two distinct behaviours (steady state and oscillatory) in parameter space. This is denoted a *supercritical* bifurcation because the cyclic oscillations occur for values of the control parameter strictly greater than the critical point. However, *subcritical* bifurcations are also possible. In this setting, there exists a region where the steady state solutions and the periodic oscillations co-exist. Outside of this zone, the system behaves in the same way as the supercritical bifurcation (i.e., a single steady state *or* a periodic pattern of activity) (**Figure 1e**). However, within this zone, the two dynamic states co-exist (**Figure 1e and f**). Noise can then push the system between these stable states, causing erratic jumps between low amplitude equilibrium and high amplitude oscillations (**Figure 1f**). This type of behaviour is called *multistability* (Freyer *et al.*, 2011; Tognoli and Kelso, 2014). We consider examples of multistability in brain and behaviour below.

Although critical fluctuations (near a supercritical bifurcation) and multistable fluctuations (due to a subcritical bifurcation) are mathematically related, they yield quite distinct statistics: The former (slow fluctuations) follow a scale-free power law distribution. This is the classic meaning of the term *criticality*. Noise-driven switches between two or more multistable attractors do *not* occur with a scale-free probability. In the setting of large additive noise, the transitions are akin to a Poisson process and thus follow an exponential distribution. With smaller, state-dependent noise, the system tends to get trapped near each state, with the transitions then following a heavier tailed stretched exponential distribution (Freyer *et al.*, 2011; Freyer *et al.*, 2012). Either way, such multistable switching does not possess scale-free (power law) properties and does not correspond to the classic notion of criticality.

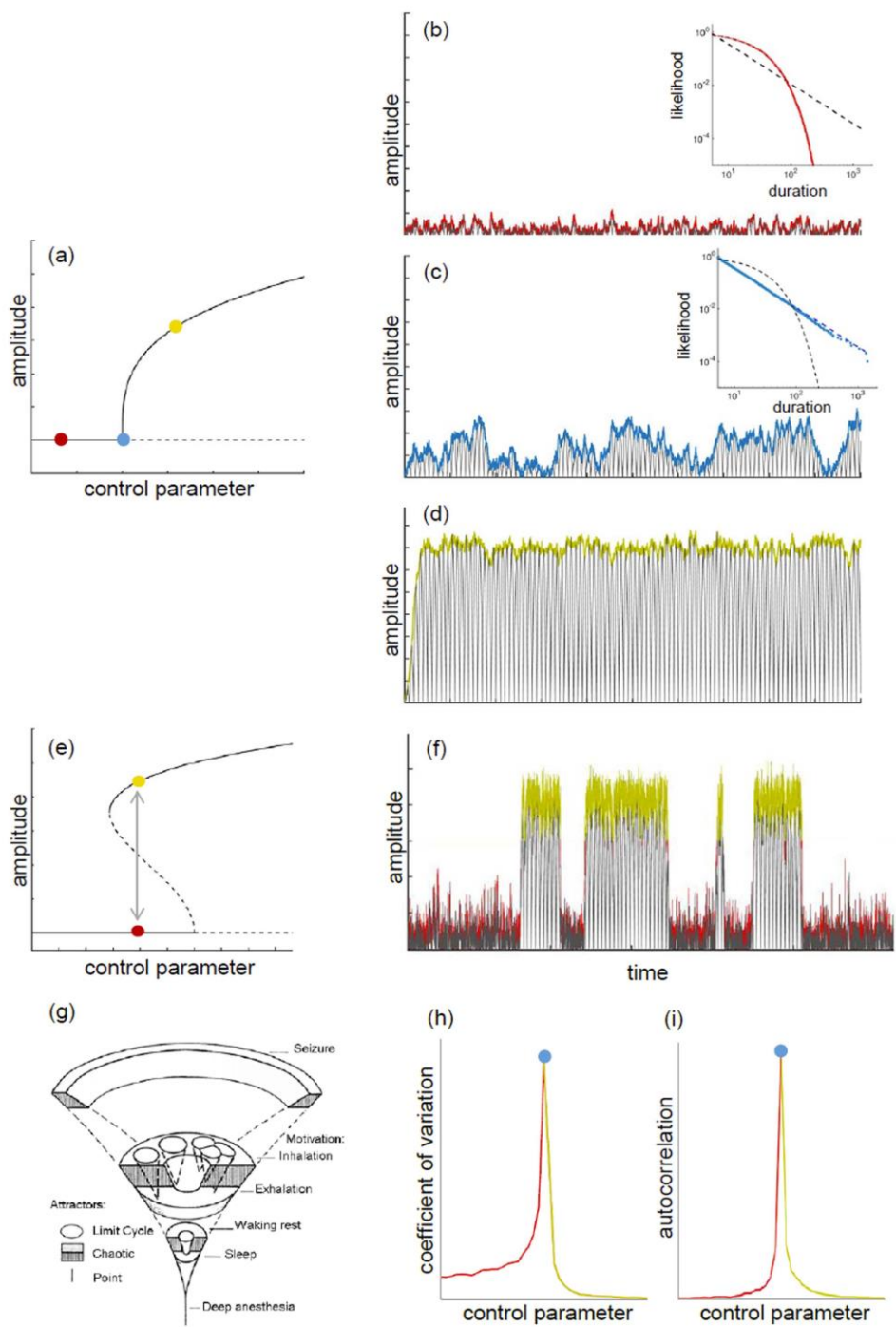

*Figure 1:* Criticality in a low dimensional system consisting of a few interacting components. (a) Super-critical bifurcation diagram, depicting the amplitude of a system's state variable (y-axis) as a function of a control parameter (such as the strength of interactions, x-axis). When the control parameter is increased, the activity of the system switches from a damped equilibrium point (red circle) to oscillatory behaviour (yellow circle). The point of change is known as the critical point (blue circle). (b) In the presence of noise, the damped system (red circle) exhibits low amplitude, rapid fluctuations. The duration of these follows an exponential probability distribution (red dots, inset). (c) At the critical point, the fluctuations have high variance and rise and fall slowly, following a power law distribution, corresponding to a linear relationship between their duration and their likelihood in double logarithmic coordinates (inset). In addition, the slope of this relation is described by a critical exponent of $\alpha = -3/2$. (d) Beyond the critical point, the system exhibits sustained oscillations. The fluctuations in the amplitude envelope of the oscillations are fast and small. (e) Sub-critical bifurcation

diagram, with a zone of co-existence (or "bistability") between the fixed point and oscillatory behaviours. In this case, system noise not only drives fluctuations around each attractor, but can also drive sudden and erratic jumps between the two dynamic states as depicted by the double headed arrow. (f) Example bistable time series. (g) Multistable bifurcation diagram as reproduced from Freeman (Freeman, 1987), proposed as a model for perceptual activity in the olfactory system. In this case, the number and complexity of the attractors is larger, however, the underlying principle is the same. Critical slowing down corresponds to a sharp increase in the coefficient of variation of the mean amplitude across 200 trials (h) and the auto-correlation function (i) at the critical point.

## 2.2. Criticality and phase transitions

We have thus far considered relatively simple systems composed of only a few components, or where the elements of the system are lumped into a small number of variables (such as all predators and all prey species each being considered a single entity). We now move to studying critical systems composed of many interacting components such as magnetic spins in iron (Stanley, 1987), grains of sand falling onto a pile (Bak *et al.*, 1988), or neurons (Plenz and Thiagarajan, 2007). On top of the example considered in **Figure 1**, these examples introduce a *spatial dimension* through which the components of the system interact.

The emergence of a magnetic field in a ferromagnetic material (such as iron) cooled below a critical temperature ($T_c$) is a classic example of a phase transition. Such materials have permanent magnetic moments (dipoles) due to the spin of unpaired electrons in atomic or molecular electron orbits. These dipoles interact through the mutual effects of the local fields that they impart on their immediate neighbours (**Figure 2a**), causing neighbouring dipoles to align and form local domains of coherent fields. These effects partially counter the influence of stochastic thermal and quantum effects that cause random flips in the direction of the dipoles. At temperatures greater than Tc, stochastic flips disrupt the formation of larger domains and, in the absence of an external field, the material will not possess a macroscopic magnetic field. Slow cooling of the material allows domains of increasing size to form, although domains at the very largest scales continue to disappear into the background noise. However, when the material is cooled to a critical point (the Curie temperature), the smaller domains coalesce into increasingly larger ones until they approach the size of the entire system. The coalescence of small domains into those of successively larger size is called an avalanche (**Figure 2a**). At the Curie temperature, avalanches have no characteristic size and thus may intermittently sweep through the entire system. These avalanches can be measured empirically using a large, external pick-up device (Cote and Meisel, 1991; McClure Jr and Schroder, 1976; Meisel and Cote, 1992; Perković *et al.*, 1995). Below the critical temperature, the mutual interactions amongst the spins align into domains that encompass nearly every dipole; at this point, the material shows a coherent ferromagnetic field (despite ongoing disorder at the atomic scales). The transition through the critical point in such a high dimensional system is called a phase transition.

A phase transition in iron cooled below its Curie temperature is a classic example of how simple internal interactions can overcome disorder and yield, through a critical point, a macroscopic field. The imposition of an external field of sufficient strength on a ferromagnet below the critical point can cause the field to suddenly switch directions to align with the applied field (Vojta *et al.*, 2013): Here, in contrast, the macroscopic order is imposed externally.

There are many similarities between bifurcations and phase transition, including the presence of super- and subcritical varieties (**Figure 2b**): These are called continuous and discontinuous or (second- and first-order) phase transitions in this context (Kim *et al.*, 1997). The transition from a para- to a ferromagnet due to cooling, outlined above, is an example of a continuous phase transition:

The sudden switching of that field due to an external field is a discontinuous one, as is water turning into vapour in the presence of heat (Stanley 1987). When the phase transition is discontinuous, then noise-driven multistability may also occur: Noise can induce switches between ordered and random states (**Figure 2b**). Alternatively, as in the case of the (discontinuous) phase transition between water and steam, there can arise complex mixtures of both. Just as in the case of a bifurcation, however, multistability arising due to a discontinuous phase transition does not exhibit the scale-free, 'critical phenomena' discussed above. Critical, power law scaling in the spatial and temporal domains is unique to a continuous phase transition.

Whilst we here focused upon the canonical example of spins in a weak external field, the basic ingredients (local interactions, noise, a large number of subsystems, an external influence that brings weak coherence) occur widely and, as a result, phase transitions are ubiquitously observed in nature (Stanley, 1987).

The temporal behaviour of a spatially-extended system near a phase transition mirrors the behaviour of two interacting elements near a bifurcation – namely slow, high amplitude fluctuations. These fluctuations, however, additionally exhibit complex spatiotemporal processes – avalanches – that also show scale-free statistical properties. Away from the critical point, the likelihood of an avalanche drops off quickly (exponentially) with size (the number of elements involved). In the classical example of magnetism, the domains of coherent spins are very small at high temperatures. If the temperature is tuned towards the critical point ($T_c$), the coherent domains sporadically increase in size and the ensuing distribution of avalanche sizes decays slowly as a function of spatial scale. The size of such domains measured over time converges toward a scale-free (power law) effect (**Figure 2c**). If the temperature is further reduced, large and stable coherent domains appear corresponding to the emergence of an internal magnetic field.

The *correlation length* is a useful concept in this setting. In the disordered phase of the system (i.e. for weak external fields) only adjacent spins are correlated– distant spins are completely uncorrelated. As larger avalanches begin to appear at low temperature, electron spins become correlated across the scale of the corresponding coherent domains. The correlation length – the spatial scale at which pairs of electrons are at least weakly correlated –increases. At the phase transition, as the size of the coherent domains approaches that of the system, the correlation length diverges upwards. As a result, external perturbations applied to any part of the system may lead to a change in the state of the whole system. Put alternatively the system has maximum dynamic range because any small perturbation has a chance of changing the electron spins of such a critical paramagnetic system.

We have focused upon phase transitions in spatially extended (embedded) systems, such as magnets and water which are dominated by interactions or collisions between the neighbouring elements of the system. However, the description of avalanches in large *N* systems does not inevitably refer to space. A *branching process* is a simple model of a phase transition that describes how activated elements may either decay (to inactive) or activate other elements with the progression of time (De Carvalho and Prado, 2000). Criticality occurs when the decay and activation rates are in equal ratio. This canonical model of a phase transition, which has been used to study criticality in neural systems [e.g. (Beggs and Plenz, 2003)], does not require a spatial dimension; metrics such as the correlation length do not make sense in these models. Although such abstract models may not take space into account, complex multi-unit physical systems such as the brain must be embedded in space, and the interactions between the elements are very often constrained by their spatial proximity (Roberts *et al.*, 2016).

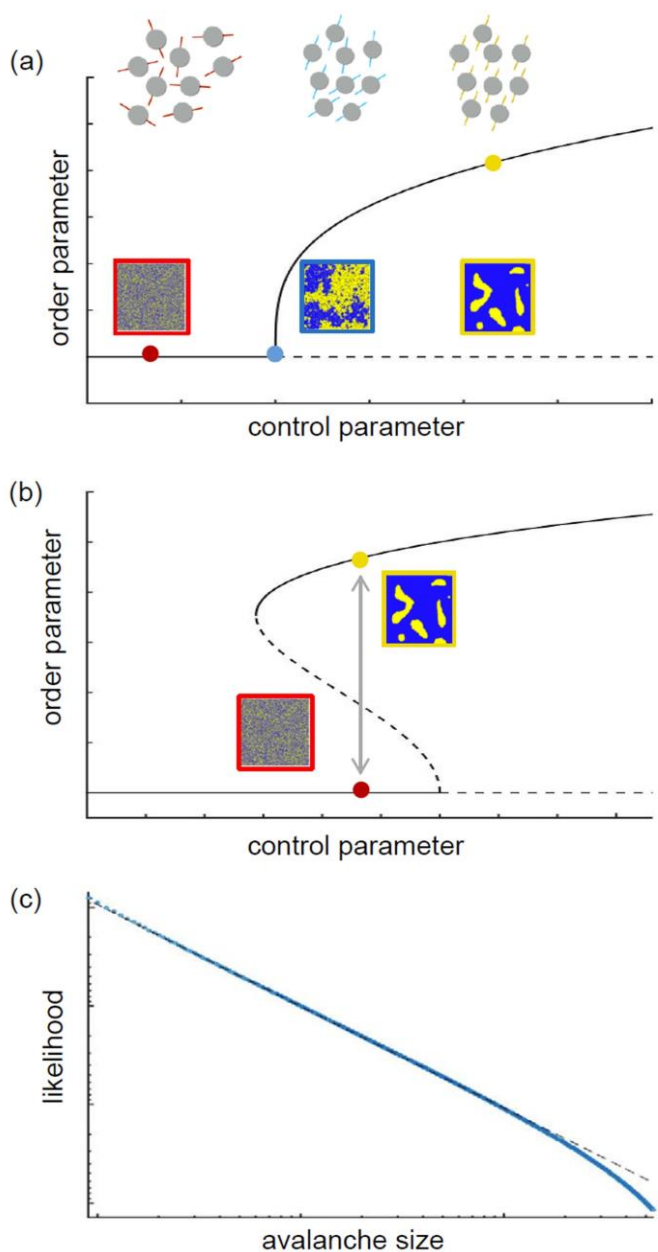

*Figure 2:* Phase transitions and avalanches in spatially extended systems. (a) Continuous (or second order) phase transition. The disordered (random) phase (red) shows lack of spatial order (red square) with randomly oriented spins in a typical physical system such as a ferromagnet in a weak external field. The ordered phase (yellow) shows large domains of co-aligned spins. At the critical point (blue), avalanches of complex partially-ordered domains rise and dissolve across all scales (blue square), leading to a power law size distribution. (b) Phase transitions can also be discontinuous (also called first order). As with a sub-critical bifurcation, system noise then causes erratic switching between the disordered and ordered phases (double headed arrow). (c) Cumulative probability distribution of the relationship between the size and likelihood of avalanches at criticality. A scale-free processes yields a linear scaling relationship in double logarithmic coordinates (a power law) with a critical exponent of $\alpha = -3/2$. Note the slight exponential truncation at the right hand side, due to finite size effects.

As we have seen, phase transitions can be considered a natural extension of the notion of a bifurcation from systems with few components, to those with many. The underlying mathematics is very similar (in fact, the so-called Landau equation, used to model generic phase transitions, is mathematically identical to the Normal form equation used to describe bifurcations). Historically, however, the two phenomena have been studied in different fields - bifurcations by applied mathematicians, whereas phase transitions were classically the domain of physicists. This legacy has led to a difference in the use of the central terms, sub- and super-criticality: In mathematics, "sub-" and "super-critical" qualify bifurcations, denoting distinct instabilities that differ in their underlying mathematical nature (**Figure 1a** and **1e**) and, as a result, the behaviour they yield. In physics, these terms are used to describe the phases of the system. "Subcritical" phase is used to denote the stable, absorbing state below a phase transition (**Figure 1a**, red dot and **Figure 1b**). The term "supercritical" is used to denote the ordered state above the transition, whereby the amplitude of the order parameter (y-axis) is typically non-zero (**Figure 1a**, yellow dot and **Figure 1d**). It is unlikely that an attempted synthesis of those terms here would pervade both fields. For the remainder of this paper, we use the terms "sub-" versus "super-critical bifurcation" to denote the type of instability, and "sub-" versus "super-critical state", "activity" or "phase" to denote where a particular system lies with respects to the phase transition. In general, since the interpretation is largely dependent on the audience, caution is required when interpreting or using those terms more broadly.

Phase transitions and criticality have been documented in a very broad range of physical systems over many decades (Kosterlitz and Thouless, 1973; Yang and Lee, 1952) and their study remains one of the most active areas of research in branches of physics such as statistical mechanics (Papanikolaou *et al.*, 2011; Sethna *et al.*, 2001; Zapperi *et al.*, 2005). The signatures of criticality have been documented in systems as diverse as flocking birds (Cavagna *et al.*, 2010), earthquakes (Burridge and Knopoff, 1967; Carlson and Langer, 1989; Rice and Ruina, 1983); solar flares (Lu *et al.*, 1993), armed conflict (Roberts and Turcotte, 1998), traffic jams (Nagel and Herrmann, 1993), and capital wealth (Roberts and Turcotte, 1998) – even crumpled paper [(Houle and Sethna, 1996; Kramer and Lobkovsky, 1996); for review see (Roberts and Turcotte, 1998)].

We turn to evidence for criticality in the brain in Section 3 after considering its underlying appeal. But before moving, it is instructive to contextualize the importance of criticality as a theoretical framework and how it may come about in many distinct systems.

2.3. The conceptual appeal of criticality

Criticality derives its basic appeal from a number of considerations. First, it speaks to the presence of a relatively simple underlying process - the response of a weakly stable system to stochastic perturbations - arising in very different settings. The processes that drive the system close to instability can be diverse – a build-up of fuel; varying temperature; a driving external magnetic field; strong interactions between species – but the collective response in generating slow, multiscale fluctuations is shared. Likewise, the many specific details of the systems differ markedly (e.g., magnetic spins, moving tectonic plates, neurons) but can be unified by their core dynamic nature - possessing interactions among their components that erratically amplify and damp microscopic perturbations. This notion of "universality" is very appealing to scientists who seek unifying principles across diverse systems. That is, the appearance of power-law and invariant scaling in markedly different systems suggests the importance of processes that transcend their particular incarnation. Of note, the characteristic exponent in the power-law scaling that describe the critical fluctuations in many of these systems typically converges to a value of $-3/2$. Theoretical considerations support the emergence of this value in systems at the cusp of a phase transition (Zapperi *et al.*, 1995). Thus, basic theoretical arguments unify diverse phenomena – this is the essence of *universality* (Stanley, 1999).

Computational considerations also underlie the appeal of criticality. Here we review the computational aspects of criticality in physical systems. In Section 3.3, we focus on the computational advantages of criticality in the brain. The earliest demonstrations of the computational advantages of the critical state were done in a very simple and idealized system called *cellular automata* (CA), whose dynamics evolve discretely in space and time according to very simple interaction rules (Langton, 1990). By changing an underlying interaction parameter, CA can be tuned to converge very quickly to simple periodic (spatiotemporal) structures, or to unstructured, chaotic processes. In between these scenarios – at the so-called "edge of chaos" – CA exhibit lengthy mixtures of ordered and disorganized structures. Theoretical arguments show that the computational complexity of CA diverge in this regime. That is, if one considers the information content of the system at each time point, the number of iterations before CA converge onto a stable solution becomes very long for this in-between state (although see (Mitchell *et al.*, 1993) for an opposing position). Depending upon one's viewpoint, the simplicity of CA is either conceptually appealing (since complexity arises from very simple laws) or distracting, because the physical meaning of CA is unclear. However, the implications of the proposal – complexity from simplicity - are tantalising, underlying the influence of CA. Its catch-phrase "edge of chaos" became a very well-known way to refer to the computational advantages offered by the critical state. Further research has shown that several other physical and biological systems also have optimal computational properties at criticality (Crutchfield and Young, 1988; Kauffman and Johnsen, 1991; Mora and Bialek, 2011; Nykter *et al.*, 2008).

The appeal of criticality also finds support from thermodynamic perspectives. In a stable, subcritical system, random fluctuations arising from thermal energy and other sources of entropy are confined to the microscopic scale, like a giant TV screen showing pixel-wise static. While these microscopic fluctuations have high *entropy*, meso- and macroscopic scales are damped and are hence in a featureless, low entropy state. Above the critical point, the macroscopic scale of the system can show interesting features, such as periodic structures and oscillations. However, fluctuations at finer scales are slaved to these large-scale features and thus do not express the potential entropy arising from the smallest microscopic scales. In these two states – sub- and super-critical, respectively, high entropy can be thought of as being trapped at one particular scale and unavailable at other scales. At the critical state, microscopic fluctuations erratically disseminate to larger scales through avalanches and crackles. These fluctuations introduce packets of disorder which accordingly increase the information content of the system across all scales. That is, while criticality increases correlations – and thus decreases entropy – at the smallest scale, it "transports" random fluctuations across scales increasing the total complexity of the system (Tononi *et al.*, 1994).

2.4. Self-organised criticality
Why is it that so many systems found in nature appear to be perched at a critical point? In theory the critical point becomes confined to a very small range of values as the size of the system increases (see **Figure 1h**). For an experimental system such as the paramagnetic material discussed above, the external field can be carefully (manually) tuned until crackling noise and avalanches appear. However, for other systems, such as earthquakes, forest fires and flocking birds the fingerprints of criticality seem to arise internally without the need for careful tuning by an external observer.

The answer to this apparent dilemma is contained in the notion of *self-organised criticality (SOC)*, a process whereby a complex system is driven toward its critical point across a very wide set of starting points and parameter values. The classic example of SOC was provided in the behaviour of sand-pile avalanches by the work of Per Bak (Bak, 1990; Bak *et al.*, 1987, 1988). In essence, the slow addition of sand to the apex of a sand-pile leads to the gradual increase in the slope of its sides. At a critical slope, scale-free avalanches of falling sand begin to occur. The slope angle decreases with each avalanche as (gravitational) energy is released from the system, then increases again as new sand is added. The slope does not need to be tuned by the experimentalist but naturally emerges from the interplay of the interactions between the adjacent grains of sand, the external (gravitational) force and the slow addition of sand.

Modelling SOC in complex systems with weak local interactions and noise is a very active field (Marković and Gros, 2014). Two mechanisms appear sufficient for the appearance of SOC – firstly the dissipation of energy and secondly some form of "memory" in the system. For example, a large forest fire burns through a build-up of timber fuel: Energy has dissipated from the system and a period of time must now pass until there is sufficient new fuel for fire of any appreciable magnitude. Likewise, a large avalanche of sand in a slowly building sand pile changes the gradient of the pile. Time and small avalanches must then accrue before the slope of the pile is sufficiently steep to trigger another large avalanche. Again in seismology, tension from tectonic plates is released following a large earthquake and its aftershocks, such that subsequent large earthquakes are unlikely to follow immediately. Each of these systems is characterized by the build-up and subsequent dissipation of energy or resources, whose release is 'remembered' by the system until sufficient resources have recovered. We will encounter similar concepts when we discuss mechanisms of SOC in neuronal systems (Section 3.3).

## 3. Criticality in the brain

The role of criticality and multistability in neurophysiological systems of the brain was first articulated over 3 decades ago by Walter Freeman following detailed empirical analyses and computational models of the rabbit olfactory bulb (**Figure 1g**). In particular, Freeman proposed that the process of inhalation and exhalation acted, via modulation of the gain of excitatory neurons, to sweep the activity of the olfactory bulb through a sub-critical bifurcation and hence through a zone of multistability (Freeman, 1987; Freeman, 1991). Sensory inputs, arising from contact of inhaled molecules with membrane receptors of olfactory neurons, then act to selectively perturb the system onto one of several competing dynamic patterns. This dynamic pattern was proposed to encode the olfactory input – the percept - until it destabilized during exhalation as the system passed again into the zone where only the stable equilibrium solution exists. We now survey more recent examples of criticality and multistability in neuronal systems that build upon the foresights contained in Freeman's prescient papers.

### 3.1. Rhythmic fluctuations, bifurcations and slowing down

The mechanisms underlying motor coordination have been an intriguing area for the application of dynamic systems theory (Bressler and Kelso, 2001; Kelso and Clark, 1982). One fruitful candidate has been the study of rhythmic finger tapping. At slow frequencies, humans are able to tap in either of two stable modes: syncopation and anti-syncopation. However, at high frequencies, the anti-syncopation mode becomes unstable and only the in-phase syncopation pattern is expressed (Kelso *et al.*, 1986). The hallmarks of criticality are seen just prior to this transition, namely slowing, high amplitude movement fluctuations (Kelso, 1984, 2014). To explain this, Haken, Kelso and Bunz used a sub-critical bifurcation in a simple model of motor coordination between the left and right motor cortices (Haken *et al.*, 1985). The transition from a bimodal to unimodal pattern of behaviour occurred at the critical value of the movement frequency. This framework has also been employed to explain the transition between movement patterns induced by transcranial magnetic stimulation (TMS) (Kelso, 2014).

More recent work on criticality has focused on the temporal fluctuations observed in the major rhythms of EEG and MEG data. Employing an analysis called *detrended fluctuation analysis (DFA)*, Linkenkaer-Hauser and colleagues reported that the fluctuating amplitudes of the two dominant oscillations of the human brain – the alpha and beta rhythms – exhibited scale-free temporal statistics (Linkenkaer-Hansen *et al.*, 2001). Fluctuating levels of synchrony between pairs of electrodes have also been reported to show scale-free statistics (Stam and de Bruin, 2004). Likewise, the power spectrum of human neocortical activity acquired from invasive ECoG data shows scale-free temporal behaviour across a very broad range of frequencies (Miller *et al.*, 2009). These frequencies also show multi-scale *nesting* – that is, the amplitude of high frequencies is coupled to the phase of lower frequencies; a pattern that is recursively repeated from very slow to very high frequencies (He *et al.*, 2010). Computational models of large-scale neuronal activity – *neural field models* – suggest that the critical temporal statistics in these electrocortical recordings may arise from a subcritical bifurcation of activity in corticothalamic loops (Freyer *et al.*, 2009; Freyer *et al.*, 2011). Such modelling proposes that noise-driven switching between a low amplitude steady state and high amplitude oscillations (**Figure 1**) yields the empirically observed critical fluctuations seen at rest (Freyer *et al.*, 2012).

3.2. Neuronal avalanches and phase transitions
In 2003, Beggs and Plenz found evidence of critical behaviour in the erratic spontaneous activity measured in *in vitro* neuronal cultures (Beggs and Plenz, 2003). They documented the two salient features of criticality in a spatially extended critical system, namely power-law scaling in time (the duration of bursts of activity) and space (the number of electrodes spanned by each burst). Together with the earlier work of Freeman, this finding ushered in criticality as a term of clear relevance to complex neuronal multi-scale phenomenon.

Since these initial reports of avalanches in *in vitro* slice preparations by Beggs and Plenz, research into critical avalanche-like activity in spatiotemporal neural recordings has proceeded at great pace (Plenz and Thiagarajan, 2007; Schuster *et al.*, 2014; Shew, 2015). Observations of scale-free spatiotemporal fluctuations in spontaneous, physiological data have progressed from *in vitro* slice preparations (Beggs and Plenz, 2003), to *in vivo* recordings from superficial layers of cortex (Gautam *et al.*, 2015; Gireesh and Plenz, 2008) to awake non-human primates (Petermann *et al.*, 2009) [for review, see Shew and Plenz (2013)]. Scale-free avalanches have been reported in human whole brain magnetoencephalographic (MEG) data (Shriki *et al.*, 2013), and complex, scale-free spatial dependences, consistent with avalanches, have been described in whole brain functional neuroimaging (fMRI) data (Tagliazucchi *et al.*, 2012). Notably, these recordings cross broad scales of aperture from multi-unit recordings to macroscopic field potentials and whole brain functional neuroimaging data.

There is a growing focus in the imaging community on spontaneous (resting-state) fMRI data and the reproducible structures these reveal in health (Damoiseaux *et al.*, 2006; Zalesky *et al.*, 2014) and disease (Fornito *et al.*, 2015). Whereas the number of channels in neurophysiological recordings, such as MEG, has a modest upper bound (of several hundred), the high spatial resolution of fMRI yields thousands of voxels (~100,000 voxels). These data hence contain the breadth of spatial scales required to interrogate whether the spatial fluctuations are scale-free (Eguiluz *et al.*, 2005) and thus whether critical dynamics underlie the dynamic patterns seen at rest (Chialvo, 2012). Recent evidence from both empirical (Tagliazucchi *et al.*, 2012) and computational (Deco and Jirsa, 2012) analyses points in favour of this proposal. Among the most intriguing findings are the recapitulation of the classic resting-state networks (Yeo *et al.*, 2011) by models of critical dynamics arising from primate (Honey *et al.*, 2007) and human (Haimovici *et al.*, 2013) structural connectomes. Analysis of the temporal statistics of resting-state fMRI and EEG also suggests that long-range, scale-free correlations may indeed lie at the very heart of the slow fluctuations that are observed in these data (Van de Ville *et al.*, 2010).

This emerging view is schematically summarized in **Figure 3**. In the sub-critical zone, bursts of cortical activity are sporadic uncoordinated (red box). Above the critical value, cortical activity is coupled too tightly and conversely, inadequately segregated (yellow box). Resting-state networks function at the critical value, where switching between network states occurs due to weak dynamic instabilities (light blue box).

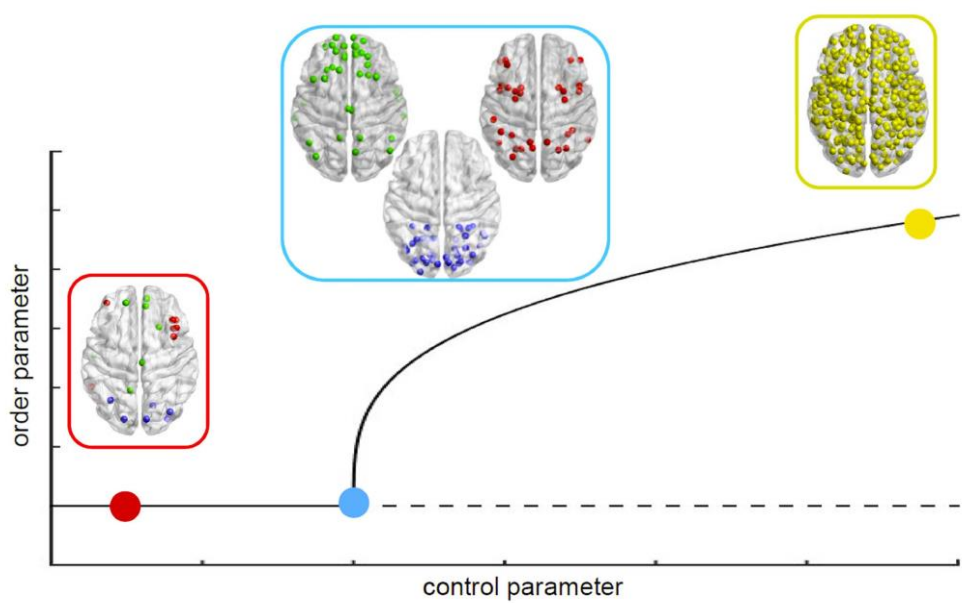

*Figure 3*: Proposed role of criticality in large-scale, resting-state brain dynamics. In the sub-critical region, individual brain regions are effectively uncoupled, showing a lack of integration (red square). Conversely, in the super-critical region, integration is too great and there is a lack of segregation (yellow). Near the critical point, an emerging body of work in EEG, MEG and fMRI suggests that brain systems show a dynamic balance of integration and segregation (blue square), fluctuating among the various resting-state networks (and EEG rhythms). Cognitive function requires a slight incursion away from the critical regime leading to a stabilization of one particular network, consistent with the earlier proposals of Freeman.

There is now a very well established relationship between resting state brain networks and the underlying structural connectome from which they arise (Honey *et al.*, 2007). Criticality arising in simple systems does not require a complex spatial substrate: Rather as we have reviewed above, its hallmark is the emergence of complex spatiotemporal processes from simple, local interactions. However, complex networks may allow critical-like behaviour to occur in a region of parameter space instead of a single point (Moretti and Muñoz, 2013). Moreover, in the primate brain, structural-functional correlations argue for the existence of a relationship between critical dynamics and the relatively static underlying structural connectome. The nature of this relationship between critical states in resting-state fMRI data and the connectome is not well understood. Highly interconnected cortical hubs, and brain regions comprising the so-called default-mode network may play a prominent role in maintaining resting-state network dynamics (Gollo *et al.*, 2015; Leech *et al.*, 2012; Vasa *et al.*, 2015) and in facilitating the efficient spread of avalanche events through macroscopic brain networks (Misic *et al.*, 2015). The constellation of densely connected hub regions – *the rich club* – appear to support a slow, stable dynamic "core" whereas peripheral sensory regions introduce (rapid) stimuli-related variability in the system (Bassett *et al.*, 2013; Gollo *et al.*, 2015; Hasson *et al.*, 2015). Such a core-periphery organization of brain network dynamics speaks to a hierarchy of time-scale fluctuations, in which hub regions integrate and regulate the network dynamics largely operating at slow frequencies (Cocchi *et al.*, 2016; Gollo *et al.*, 2017; Gollo *et al.*, 2015; Hasson *et al.*, 2008; Honey *et al.*, 2012; Murray *et al.*, 2014). Hub regions within the default-mode brain network may represent a structural signature of near-critical behaviour. Regions comprising this network exhibit coordinated activity in the resting-state when the coherence between nodes of other "task-positive" networks is generally suppressed (Cocchi *et al.*, 2013; Fox *et al.*, 2005; Hearne *et al.*, 2015).

These observations suggest mechanisms through which critical dynamics may adapt to, and reshape the complex nervous systems in which they occur, particularly the relationship between synaptic processes, functional connectivity and network topology (Rubinov *et al.*, 2009; Zhigalov *et al.*, 2017).

### 3.3. Computational aspects of neuronal criticality

The pioneering work of Freeman (on bifurcations and multistability), and Beggs and Plenz (on critical avalanches) provided proof-of-principles that the science of criticality could be used to inform our understanding of complex patterns of activity in the brain and, by extension, behaviour. Research in these areas has accelerated dramatically and now yields a stream of important discoveries spanning from the neuronal (Gal and Marom, 2013; Gollo *et al.*, 2013) to the whole-brain scale (Kitzbichler *et al.*, 2009).

As reviewed above, the study of criticality in physical systems using simple models suggested that systems at the critical state are endowed with optimal computational properties. Such advantages have recently been demonstrated in models and empirical recordings of critical neuronal systems [for review, see (Beggs, 2007; Shew and Plenz, 2013)]. Perhaps most crucially, optimal dynamic range – the sensitivity of a neuronal system to respond to, and amplify, inputs across a broad spectrum of intensities – was shown to be maximized in models of neuronal systems tuned to a critical state (Kinouchi and Copelli, 2006; Larremore *et al.*, 2011). In slice cultures grown on the surface of multielectrode arrays, Shew and colleagues later provided empirical evidence for this proposal, showing that the maximum dynamic range to electrical perturbation was indeed maximized when the cultures were pharmacologically manipulated to be close to criticality (Shew *et al.*, 2009). Recent electrophysiological recordings from the anaesthetized rat provided the first *in vivo* evidence that dynamic range in perceptual systems is maximized when background activity is at the critical point (Gautam *et al.*, 2015).

Another example of the computational advantages of criticality arises from simplified neuronal models which predict high fidelity and optimal information transmission (maximum mutual information between sender and receiver) at criticality (Beggs and Plenz, 2003; Greenfield and Lecar, 2001): This prediction was also later observed *in vitro* (Shew *et al.*, 2011) and more recently in awake, behaving mice (Fagerholm *et al.*, 2016).

Research in this field has also suggested that information storage and capacity – the ability of a system to encode a broad repertoire of complex states from which information can be decoded (Gatlin, 1972)– may be optimized at criticality. Again, this notion has been captured in simple neural models (Bertschinger and Natschläger, 2004; Haldeman and Beggs, 2005; Yang *et al.*, 2017) and also demonstrated in empirical recordings, including those arising in unperturbed (resting state) recordings (Breakspear, 2001; Deco and Jirsa, 2012) as well as through careful pharmacological manipulations of *in vivo* neurophysiological recordings (Stewart and Plenz, 2006). Selective enhancement of weak (but not strong) stimuli also occurs near a phase transition (Copelli, 2014). It is also interesting to note that optimal computational properties can arise at both continuous and discontinuous phase transitions (Gollo *et al.*, 2012).

A compelling argument for the advantage provided by the critical state derives from the analysis of psychophysics experiments, which quantify the relationship between physical stimuli and perceptual responses. Psychophysics relations may represent the earliest documented evidence of critical dynamics in the nervous system (Kello *et al.*, 2010). Across a variety of sensory modalities, these experiments showed power-law relations, known as Steven's laws, in which the perceived psychophysical or neuronal response $F$ is given by: $F(S) \propto S^m$, where $S$ is the stimulus level and $m$ is the Stevens exponent (Stevens, 1975). To account for the saturation of the response that occurs for extreme stimuli, the psychophysics response is modelled by a sigmoid Hill function: $F(S) \propto F_{max} S^m / (S^m + S_0^m)$, where $F_{max}$ corresponds to the saturated response, and $S_0$ the input level for half-maximum response. Importantly, this function also retains the power-law regime governed by the exponent *m*. This power-law behaviour has a key putative function: It allows animals to distinguish stimulus intensity varying across many orders of magnitude. As such, the power-law regime can compress decades of stimulus variation $S$ into a single decade of response $F$. A standard means to measure this coding performance is called the *dynamic range*. The larger the dynamic range, the better the ability to detect changes in stimuli. Modelling the neuronal behaviour at the sensory periphery, Kinouchi and Copelli showed that a large dynamic range emerges from the collective response of a network of many interacting units, and, more importantly, the dynamic range is optimal when the network is at the critical state (Kinouchi and Copelli, 2006). This work explains the long-lasting psychophysical scaling relations (Steven's law), and provides a clear example in which a meaningful biological feature is optimised at criticality. Similar to earlier proposals, the work of Kinouchi and Copelli was also based on a fairly simple model of neuronal activity. Crucially, however, the model generated a prediction that was subsequently verified experimentally (Shew *et al.*, 2009).

The explanation of the power-law regime of psychophysics laws in terms of the optimal sensitivity of critical states was an important contribution. However, as usual in analytic approaches, some simplifications were made, leading to at least two conundrums. The first challenge refers to the important trade-off between sensitivity and specificity. The finding that optimal signal coding occurs at criticality implies maximum sensitivity. Yet, the specificity of this state is compromised because of increased levels of fluctuation. As illustrated in **Figures 1** and **2**, criticality corresponds to the state with the largest macroscopic fluctuations. This is why the critical state allows for the amplification of stimuli of small intensity, which enhances the ability to distinguish the stimulus intensity varying over orders of magnitude (i.e., large sensitivity). However, the very same effect is also a limitation because the high fluctuations of criticality reduce the specificity of the system. The issue is whether the improved sensitivity remains beneficial to the system when the reduction in specificity is also taken into account. Fortunately, a solution for the sensitivity-specificity conflict exists in the presence of diversity amongst components of the system. Heterogeneous excitable systems exhibit recruitment properties in which units are recruited following their order of excitability (Gollo *et al.*, 2016). Therefore, a state of optimal sensitivity has units forming a subpopulation in a critical regime as well as subpopulations operating in a non-critical state with improved reliability. Hence, optimised systems represent the coexistence of subpopulations of reliable units and poor sensitivity (poised away from their critical state) with a subpopulation of unreliable units and great sensitivity (typical of the critical state). In other words, optimal perceptual performance may rely on the contribution of critical and non-critical units (Gollo, 2017).

The second challenge emerging from the explanation of the power-law regime of psychophysics laws derives from to the fact that the critical state for optimal psychophysical response of Kinouchi and Copelli separates an active state from an inactive state (Kinouchi and Copelli, 2006). The issue here is that the inactive state corresponds to an absorbing state in which the system cannot escape unless external stimuli are provided. Hence, if the system falls into this state, it will get trapped there. Critical systems typically do fall into this state because of their enhanced fluctuations. If the system corresponds to the brain or a part of the nervous system, such an inactive state would be expected to occur rather frequently. However, such silent states are not observed *in vivo*. The solution to this issue also derives from the incorporation of an essential, but often overlooked, ingredient: inhibition. Larremore and colleagues showed that ceaseless activity and critical avalanches coexist when a substantial fraction of the units are inhibitory (Larremore *et al.*, 2014). Hence, evidence is gradually accumulating for the previously idealised proposal that biological systems can exploit special features of the critical state. Meanwhile, models supporting this proposal are incrementally incorporating greater physiological detail.

Research in the auditory system has highlighted the advantages conferred by active responses of hair bundles poised in a critical state (Camalet *et al.*, 2000; Eguíluz *et al.*, 2000). In response to sound-wave input, hair bundles oscillate and their dynamics vary from a steady to an oscillatory regime. Each bundle is tuned to a particular natural frequency and adjacent cells respond maximally to successive pitches, giving rise to a tonotopic map in the cochlea (Romani *et al.*, 1982). Crucially, computational modelling suggests that active and nonlinear hair bundles that operate near a Hopf bifurcation optimise the cochlea's performance and enhance the main features of auditory coding, such as, amplification, frequency selectivity and compressive nonlinearity (Hudspeth, 2008; Maoiléidigh *et al.*, 2012).

3.4. Self-organised neuronal criticality
A growing body of empirical work has thus asserted the presence of scale-free statistics across a diversity of *in vitro* and *in vivo* neural recordings, while computational models have highlighted its computational advantages. Of note, the critical activity observed by Beggs and Plenz (2003) was stable for many hours and did not require careful tuning of the parameters of their culture (i.e. the pH, temperature, etc). Simple models of criticality classically rely upon fine-tuning of system parameters to a critical value (Levina *et al.*, 2014). What is the basis for a robustness that apparently eschews the need for such a balancing act? As discussed above (Section 2.4), analyses of critical systems in physical systems reconcile this paradox by recourse to self-organised criticality (SOC). In brief, SOC arises when the interactions amongst system components are imbued with some form of plasticity, such as when system energy accumulates and is then dissipated by a large-scale avalanche. During periods of quiescence, energy slowly accumulates until it tips the system into (or above) criticality. The consequent energy dissipation briefly renders the system sub-critical until further energy accumulates. There is thus a *time scale separation* between the fast system dynamics and the slow build up and dissipation of energy.

Several models of criticality in the brain incorporate such slow processes (Marković and Gros, 2014). A considerable body of research has focused upon the role of various forms of synaptic plasticity, including simple activity-dependent up- and down-regulation (de Arcangelis, 2008), activity-dependent synaptic plasticity (de Arcangelis *et al.*, 2006), synaptic potentiation (Stepp *et al.*, 2015), short-term synaptic depression through depletion of synaptic vesicles (Bonachela *et al.*, 2010; Levina *et al.*, 2014; Mihalas *et al.*, 2014; Millman *et al.*, 2010), Hebbian (Van Kessenich *et al.*, 2016) and anti-Hebbian synaptic plasticity (Cowan *et al.*, 2014; Magnasco *et al.*, 2009), and spike-time dependent plasticity (de Andrade Costa *et al.*, 2015; Rubinov *et al.*, 2011). As with physical systems, the (relatively) slow synaptic plasticity serves to broaden the critical point to a broad, stable region. Other neurobiological processes have also been proposed, including balanced excitation-inhibition and network topology (Rubinov *et al.*, 2011) and dynamic neuronal gain (Brochini *et al.*, 2016). More recently, the role of energy build-up and dissipation in physical systems has been recast in critical neural systems as the replenishment and depletion of intracellular metabolic resources including Adenosine Triphosphate (ATP); (Roberts *et al.*, 2014b; Stramaglia *et al.*, 2015; Virkar *et al.*, 2016).

## 4. Challenges and pitfalls of the criticality hypothesis

Despite this recent emergence of criticality research in neuroscience, lessons learned in other branches of science raise important pitfalls and caveats. First, inferring the presence of scale-free statistics in neuroscience data has classically rested upon fitting a power-law (or *Pareto*) regression to the probability distribution of the size of the temporal or spatial fluctuations (**Figure 1** and **Figure 2**). The statistical principles underlying this exercise were critiqued in a highly influential survey by Clauset and colleagues (Clauset *et al.*, 2009). While a linear regression in double logarithmic coordinates can yield a fit that looks impressive, such a process is insensitive to the distribution of data at the right-hand tail of the distribution – the very region where the presence of a heavy-tailed power law needs to be rigorously tested. This is because the number of empirically measured samples found in the tail of the distribution is often too limited for robust inference. Another concern is that the samples of a cumulative distribution function are not independent, whereas regression assumes data independence. Clauset and colleagues developed a more principled approach based on maximum likelihood estimation to test and compare different statistical models of the data, including the power law, but also other candidate heavy tailed distributions including the log normal and stretched exponential forms (Clauset *et al.*, 2009; Vuong, 1989). In brief, once the best fitting power-law parameters for an empirical dataset have been determined with maximum likelihood estimation, the goodness-of-fit between the fitted power law and empirical distribution is tested with the *Kolmogorov-Smirnov* (K-S) statistic. To this end, Clauset and colleagues proposed to randomly sample data from the fitted power law, independently fit a new power law to each of these new data samples and then evaluate the goodness-of-fit between the new samples and the new power laws. This is repeated many times to generate an empirical distribution of K-S statistics, which can then be used to compute a *p*-value for the K-S statistic corresponding to the observed data. If this *p*-value is significant, the randomly sampled data is a better fit to the power law than the observed data, and thus a power law should be excluded as an appropriate model. Otherwise, if the *p*-value is not significant, power-law behaviour is supported and the final step is to exclude other distributions as providing better evidence. Relative fits are typically computed between candidate distributions (lognormal, stretched exponentials, etc.) using log-likelihoods.

Using this approach, Clauset et al. (2009) revisited several physical phenomena thought to have scale-free statistics and showed that several of these data were better explained by other long-tailed distributions, *not* power laws. Using these methods shows that the same holds true for many neuronal fluctuations (Roberts *et al.*, 2014a). For example, it appears that fluctuating alpha rhythm follows a stretched exponential distribution, not a power law (Freyer *et al.*, 2009). In turn, biophysical models suggest that the alpha rhythm arises from noise-driven multistability, rather than being generated by classic (super-) criticality (Freyer *et al.*, 2011).

A second caveat was issued by the empirical analyses and modelling work of Tomboul and Destexhe (Touboul and Destexhe, 2010, 2015) who showed that under certain situations, the aggregate behaviour of non-critical stochastic systems could yield irregular time series with power law statistics, albeit over a limited range. This is an important issue which also highlights the importance of null models for the different experimental methods (Farmer, 2015). These findings suggest that inferences regarding criticality based on the observation of power law scaling in empirical data should be made with caution, particularly if the scaling extends for less than two orders of magnitude or the slope of the power law is steep (Miller *et al.*, 2009) - i.e. the scaling exponent α is greater than 2.5.

A third caveat concerns other classes of interesting, emergent phenomena. Non-trivial, emergent dynamics can arise through other complex nonlinear phenomena. A classic example is that of so-called winnerless competition (Melbourne *et al.*, 1989; Rabinovich *et al.*, 2001) that has been proposed to underlie cognitive tasks such as animal gait (Golubitsky *et al.*, 1999), perceptual rivalry (Ashwin and Lavric, 2010), and sequential decision-making (Rabinovich *et al.*, 2008). Winnerless competition arises from *metastable* transitions along a sequence of unstable states (not unlike the drawings of M.C. Esher). Unlike criticality, the successive states are not weakly stable, but are unstable (**Figure 4**). In common with critical systems, noise plays a crucial role in a metastable system. However, the statistics of a metastable system are *not* power laws (**Figure 4b**). Rather the duration that the system dwells near each of its states varies in proportion to the logarithm of the noise amplitude (Ashwin *et al.*, 2006). The temporal statistics therefore have a characteristic (and relatively short) time scale, and are not scale-free (**Figure 4b**). While multistable (**Figure 4a**) and metastable (**Figure 4b**) systems are therefore related (particularly in name!), their statistics are distinct and the underlying nonlinear causes differ. Unfortunately, the two terms are often used interchangeably.

A number of final caveats pertain to the practical aspects of empirical data. First, the algorithm developed by Clauset et al. (2009) assumes that there is no upper bound to the empirical power-law distribution. This is a flawed assumption for most experimental data, which inevitably derive from a finite number of sensors, and may bias model selection. Recent work has revisited this assumption, developing methods that test the likelihood of a power law with a simple cut-off (Langlois *et al.*, 2014; Shew *et al.*, 2015). Second, as mentioned above, the range of many data tested for power law scaling often span less than two orders of magnitude, yielding data that is particularly sparse in the right-hand tail (precisely where power law scaling is most clearly expressed). Although the use of model estimation rather than linear regression partly mitigates this, disambiguating amongst the variety of candidate heavy-tailed distributions can only be reliably performed when the data scale over more than two orders of magnitude. Lengthy acquisitions may help here. For example, free-living activity patterns in humans derived from accelerometry recordings over seven consecutive days scale across four orders of magnitude: These allow for disambiguation of composite exponential and truncated power law distributions in active versus inactive periods of the day (Chapman *et al.*, 2016).

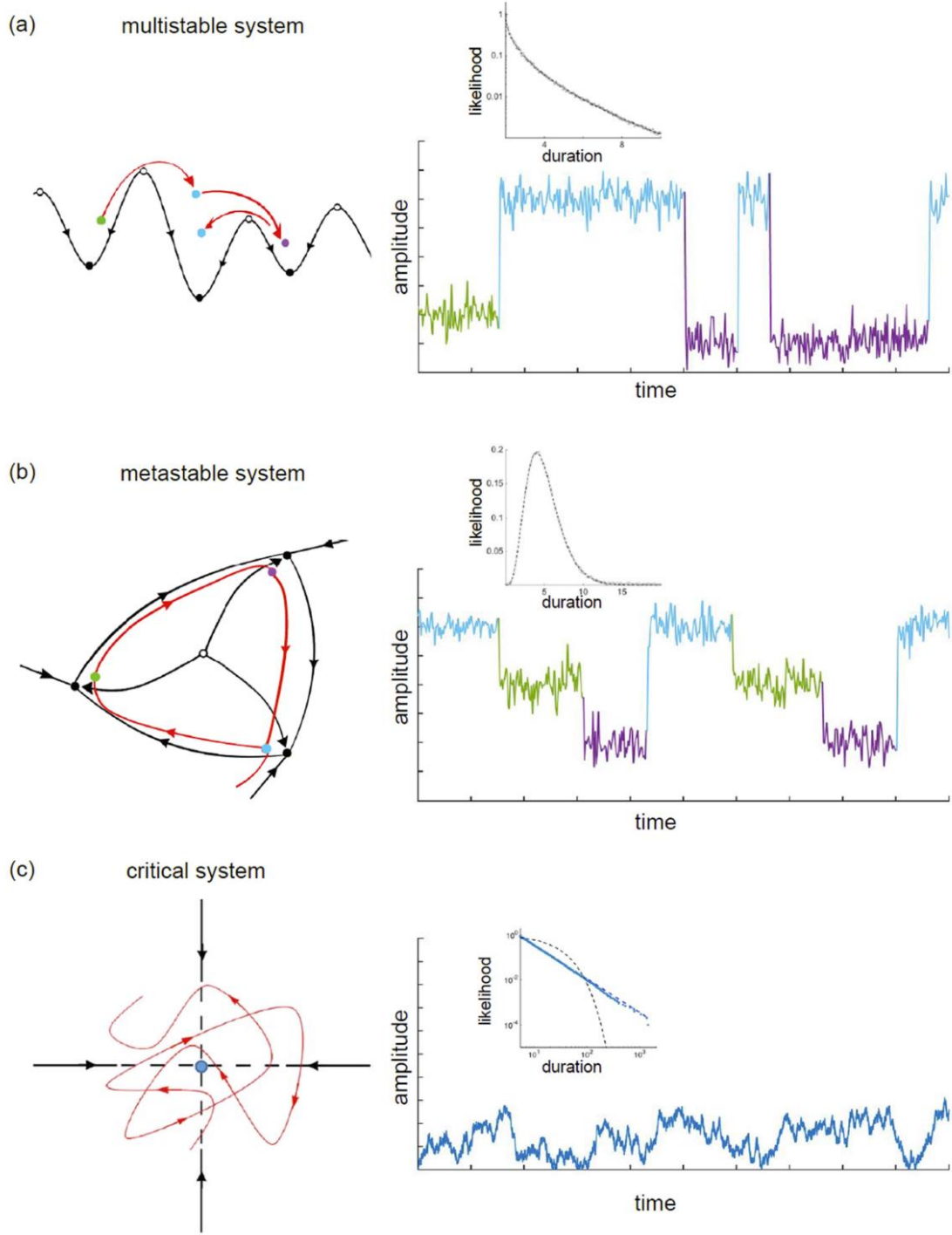

*Figure 4:* Different expressions of instability lead to different types of complex dynamics. (a) In a multistable system, noise drives a system erratically between different attractors. Because the system is briefly trapped in each basin of attraction, the time series shows a relatively long-tailed (stretched exponential) dwell distribution, here shown in linear-log coordinates (inset). (b) In a metastable system, there are no attractors, but rather a sequence of linked unstable fixed points. Because these are only weakly unstable, the system dwells in the neighbourhood of each, but does not show trapping. The sequential dwell times are therefore not long-tailed but show a characteristic time-scale corresponding to the peak in a gamma function shown here in linear coordinates (inset). In a critical system, a single fixed point is very weakly attracting or neutral. System noise leads to long and unstructured excursions corresponding to scale-free fluctuations and corresponding power-law statistics. Disambiguating these different underlying causes of complex dynamics

can be achieved with careful analyses of the system statistics, together with inversion of corresponding computational models.

Finally, noisy fluctuations in physiological data do not only arise from the system of interest (i.e. neural activity), but also from the imperfect measurement process: In the setting of neurophysiological data, these fluctuations consist of additive noise from physiological sources (such as muscular activity, cardiovascular influences) as well as artefacts due to extraneous effects (e.g. thermal scanner noise, head motion). In principle, such effects could lead to false positives in power-law evaluation. However, these inputs are generally uncorrelated and their summation therefore (according to the central limit theorem) likely to be Gaussian, not heavy-tailed. Nonetheless, care should be taken to disambiguate their contribution to any putative heavy-tailed system statistics in case one particular artefact (such as head movement in the scanner) dominates. Methods of doing this include: (i) taking independent measurements of these artefacts, such as taking empty scanner room recordings (Shriki *et al.*, 2013) and ensuring that they do not possess the same statistics as attributed to the underlying neuronal system (Kitzbichler *et al.*, 2009); (ii) using algorithms such as independent components analysis or source reconstruction to unmix neuronal fluctuations from physiological and measurement noise (Freyer *et al.*, 2009); (iii) and using a formal inversion framework that formally accommodates measurements effects including spatial or temporal filtering and additive noise (Razi *et al.*, 2015).

These caveats highlight crucial points. While the application of criticality to neuroscience is an exciting field, progress needs to proceed with due caution. Analyses of neuroscience data for power laws first needs to consider other heavy-tailed candidate distributions. Second, inference should ultimately be based upon models of the causes of the observed statistics and avoid a direct inference of criticality that is based only upon data analysis. Third, computational models of neural systems that are based upon criticality should be tested closely against empirical data using appropriate frameworks (Daunizeau *et al.*, 2009; Penny, 2012) that allow disambiguation against competing models that invoke other nonlinear mechanisms [for review, see Roberts *et al.* (2015)].

While these caveats highlight important limitations, experimental manipulations and recent theoretical developments offer new opportunities to explore the "criticality hypothesis". As noted at the outset of this review, the notion of universality (properties that transcend the details of a particular system and are thus found in many diverse settings) is one of the central appeals of criticality. Universal scaling laws - such as the presence of a scaling function that inter-relates the common underlying shape of critical fluctuations across temporal and spatial scales (Sethna *et al.*, 2001; Zapperi *et al.*, 2005) - can be extracted from data and subject to null hypothesis testing (Friedman *et al.*, 2012; Roberts *et al.*, 2014a). Relationships between the exponents of different (spatial and temporal) scaling laws, may also be derived from empirical data (Friedman *et al.*, 2012; Roberts *et al.*, 2014a) and benchmarked against the simple relationships predicted by the mathematical theory of phase transitions (Sethna *et al.*, 2001). Also, as reviewed above, the theoretical advantages of criticality in neuronal models was demonstrated in a series of elegant empirical studies using pharmacological manipulation to sweep systems from subcritical to critical to supercritical (Gautam *et al.*, 2015; Shew *et al.*, 2011; Shew *et al.*, 2009; Tagliazucchi *et al.*, 2016): Showing a sudden change in a scaling law, corresponding to a peak in information capacity or dynamic range, provides convergent evidence for the occurrence of a phase transition in the underlying system.

## 5. Emerging role of criticality in cognition

Notwithstanding the aforementioned caveats, growing empirical and modelling research clearly supports the view that neural dynamics likely occur near critical instabilities. The recognition of the limitations of this new field simply shows that it has matured beyond the "proof of principle" stage (Feyerabend, 1993). The scene is thus set for the translation of criticality into cognitive and clinical brain research.

In Section 3.1, we noted a canonical example of critical fluctuations near the transition from anti-syncopated to syncopated rhythmic finger tapping. Do critical dynamics generalize to other behaviours? Accelerometer-based analyses of free behaviour in humans (going about their everyday lives) shows that periods of inactivity exhibit power-law statistics (Nakamura *et al.*, 2007), whose scaling coefficients differ between wake and sleep (Chapman *et al.*, 2016) and differ again in major depression (Nakamura *et al.*, 2008). Intriguingly, accumulating evidence suggests that periods of activity – although long-tailed – do not fit a power law, but rather a stretched exponential (e.g., inset **Figure 4a,** depicting the Weibull distribution). This shift in the statistical proprieties of a system as a function of context is emerging as a powerful tool to understand the neural principles supporting healthy and pathological brain functions.

### 5.1. Criticality in brain and behaviour

In traditional cognitive neuroscience experiments, separate trials are typically treated as independent. However, they are not necessarily treated independently by research participants. Indeed it has been shown that reaction times between sequential trials show long-range correlations across a wide diversity of tasks (Palva and Palva, 2011; Thornton and Gilden, 2005). This is perhaps not surprising, given that, outside the scanner, human activity (Sreekumar *et al.*, 2016; Sreekumar *et al.*, 2014) and memory (Nielson *et al.*, 2015) show a complex temporal structure that scales across many orders of magnitude. Detailed analyses of reaction time data in typical psychophysics experiments support the presence of a power law structure (Van Orden *et al.*, 2005). However, definitive support for this position, as well as putative underlying causes such as criticality, do remain contentious (Heathcote *et al.*, 2000; Heathcote *et al.*, 1991; Wagenmakers *et al.*, 2004).

A rich literature on critical dynamics in brain and behaviour exists. However, with few notable exceptions (Jirsa *et al.*, 1994; Kelso *et al.*, 1992), these two streams of research have preceded largely in parallel. Several recent papers have stepped toward a unifying framework. By acquiring high density MEG/EEG data while participants performed an audiovisual threshold-stimulus detection task Palva et al. (Palva *et al.*, 2013) showed that the critical exponents of scale-free neuronal dynamics correlate with the inter-individual variability in behavioural scaling laws. These correlations, show a specific anatomical pattern, with the combined delta and alpha frequency bands correlations mapping onto posterior parietal cortex. The same cortical areas appear to be involved when electrophysiological signals in the beta and gamma ranges are considered, with the addition of the cuneus and inferotemporal brain areas. Interestingly, a significant association between neural and behavioural long-range temporal correlations was also found in brain regions comprising the default mode brain network (Greicius *et al.*, 2003). Notably, the brain-behaviour association found in task execution largely overlaps with an association between visual behavioural performances and scale-free dynamics observed in resting-state MEG data. This suggests that the temporal structure of endogenous neural dynamics are, to some extent, preserved during task performance. Similarly, the analysis of auditory task data showing that cortical activations organize as neural avalanches in both visual cognitive and resting-state contexts [(Arviv *et al.*, 2015), see also (Shew *et al.*, 2015)].

Moreover, near-critical neural dynamics may determine fluctuations in perceptual and cognitive processes. Overall, these findings highlight the strong interdependence between the near-critical neural processes characterizing resting-state and task-specific processes.

5.2. Suppression of criticality during task performance

Using simultaneously acquired EEG and fMRI data, Fagerholm et al (Fagerholm *et al.*, 2015) recently recapitulated previous findings of scale-free cortical dynamics at rest [e.g., (Palva *et al.*, 2013)]. However, as participants engaged in tasks of increasing attentional load, the statistics of cortical activity appeared to shift increasingly further from the critical state. It was surmised that although criticality was important for the unconstrained "exploratory" resting-state, executing specific tasks required suppression of the associated variability under the influence of dorsal attentional networks (Fagerholm *et al.*, 2015; Hellyer *et al.*, 2014). In keeping with this notion, complementary computational analyses have suggested that altered anatomical connectivity, cognitive flexibility and information processing following traumatic brain injury are linked to a reduction in the variance of functional connectivity simulated by a model of coupled phase oscillators (Hellyer *et al.*, 2014).

The proposition that near-critical neural dynamics are central to the emergence of conscious cognition is consistent with the results of high-density electrocorticography recordings (ECoG) in primates, showing that loss of consciousness is characterized by a reduction in the number of eigenmodes (the number of "excited modes" in a dynamic system) that are close to instability (Solovey *et al.*, 2015). Conversely, the return of consciousness appears to be accompanied by a corresponding increase in the number of eigenmodes close to instability. Recent detailed analyses of *in vivo* electrophysiological and two-photon recordings in rodents provide direct support for a clear association between the appearance of critical activity and the emergence of consciousness from anaesthesia (Bellay *et al.*, 2015; Scott *et al.*, 2014). Likewise, induction of unconsciousness through propofol anaeasthesia is accompanied by a loss of the signatures of criticality in fMRI data (Tagliazucchi *et al.*, 2016).

Non-invasive and invasive human neuroimaging work assessing neural critical states by the mean of neural avalanches analysis further support the link between near-critical neural regimes and behaviour (Palva *et al.*, 2013; Priesemann *et al.*, 2013; Tagliazucchi *et al.*, 2012). For example, ECoG data collected from patients with refractory partial epilepsy were analysed as a function of vigilance (Priesemann *et al.*, 2013). In general, the probability distribution frequency of neural avalanches followed a power law. However, slow wave sleep (SWS) was characterized by the most frequent occurrence of avalanches; unconstrained wakefulness showed an intermediate occurrence, while rapid eye movement sleep (REM) showed the fewest. These results are in line with previous findings from invasive recording in rats (Ribeiro *et al.*, 2010) and highlight the close relationship between changes in near-critical dynamics and distinct mental states.

It has been suggested that near-critical dynamics in the default mode operate to increase the "dynamic repertoire" of the brain when subjects are at rest (Deco and Jirsa, 2012; Deco *et al.*, 2011) – that is, to increase the number of proximal "cognitive sets" or stored memories (Scarpetta and de Candia, 2013). This is consistent with modelling and empirical work, reviewed above, that reveals that criticality optimizes the entropy (Fagerholm *et al.*, 2016) and information capacity (Shew *et al.*, 2011) of spontaneous neuronal activity. Criticality at rest could also function to tune cortical dynamics to an optimal state of "expected uncertainty" concerning the nature of as yet unencountered sensory inputs (Friston *et al.*, 2012b). That is, by facilitating a flexible, but also constrained degree of instability in sensory-perceptual systems, criticality could endow the cortex with the ability to adapt to volatile and non-stationary environmental fluctuations: Away from the critical point, internal fluctuations are stable and strongly damped, hence shrinking toward a baseline (mean) value. Close to the critical point, such fluctuations encompass a broad spectrum of scales, similar to the scale-invariant statistics of natural scenes (Field, 1987; Ruderman and Bialek, 1994). At a very fundamental level, this matching of internally-generated fluctuations to the statistics of external scenes, via criticality, revisits the notion that spontaneous cortical activity encodes an optimal model of the environment (Berkes *et al.*, 2011) and uses this to deploy adaptive behavioural strategies (Fiser *et al.*, 2010), including saccadic (Friston *et al.*, 2012a) and heavy-tailed fixational eye movements (Roberts *et al.*, 2013). Upon task execution - as uncertainty decreases - this unstable dynamic landscape could accordingly be suppressed, allowing stabilization of a single task-related attractor. Such proposals are consistent with the broader recognition that "brain noise" plays an adaptive role in health (Garrett *et al.*, 2011) and ageing (McIntosh *et al.*, 2010), but decreases during task execution (Churchland *et al.*, 2010; Ponce-Alvarez *et al.*, 2015).

By analysing the statistics and spatiotemporal scaling laws of neural avalanches from *ex vivo* local field potential (LFP) recordings of visual cortex, Shew *et al.* (2015) showed that strong visual stimulation initially engendered super-critical dynamics. However, with the continuation of the visual input, these were quickly tuned to a critical state through neuronal adaptation. A simple neuronal model with the adaptation to input mediated by short-term synaptic depression was able to capture the switch from non-critical to critical activity with continuous input (Shew *et al.*, 2015). These results highlight the importance of synaptic plasticity in switching between critical and non-critical regimes in order to facilitate perception and cognition.

## 6. Criticality in disease
### 6.1. Bifurcations and seizures

Whereas the role of criticality in cognition is relatively nascent, casting seizures as dynamic disorders that arise out of critical instabilities is supported by an appreciable body of evidence (Da Silva *et al.*, 2003; Meisel *et al.*, 2012). The primary generalized seizures of childhood - *Absence seizures* – correspond to the presence of high amplitude 3 Hz spike-and-wave oscillations that appear and terminate equally quickly. These seizures have been modelled as critical bifurcations in corticothalamic loops by several groups (Destexhe and Sejnowski, 2001) although there is yet a lack of consensus as to whether these bifurcations are subcritical (Fröhlich *et al.*, 2010; Suffczynski *et al.*, 2004) or supercritical (Breakspear *et al.*, 2006; Robinson *et al.*, 2002). Generalised tonic-clonic seizures, which are associated with the progression of various high amplitude waveforms, have been modelled as a subcritical bifurcation (Breakspear *et al.*, 2006). Accordingly, patients with epilepsy "reside" in pathological multistable dynamic regimes, due to neurophysiological disturbances and are occasionally perturbed into seizure dynamics (Breakspear *et al.*, 2006). Fast epileptic activity recorded invasively with cortical surface recordings have also been subject to detailed models using critical bifurcations (Bartolomei *et al.*, 2001; Wendling *et al.*, 2005). These are proposed to reflect aberrant interactions between excitatory and inhibitory cortical neurons due to GABA-ergic dysfunction (Wendling *et al.*, 2002). A detailed account of the complex progression of low frequency (DC-like) shifts in the electrical baseline and the nested high frequency oscillations that superimpose on these has recently been advanced using bifurcations: This "epileptor" - a set of equations that express the corresponding nested bifurcations - represents a comprehensive integration of slow and fast time scales underlying criticality and bifurcations in seizure activity (Jirsa *et al.*, 2017; Jirsa *et al.*, 2014).

## 6.2. Crackling noise and neonatal burst-suppression

Burst suppression (BS) is a class of electrocortical activity that occurs in preterm neonates, full-term newborn infants with encephalopathy, and following propofol anaesthesia. BS is characterised by high amplitude, irregular bursts that erratically punctuate a flat EEG trace (Niedermeyer *et al.*, 1999) (**Figure 5**). Analyses of BS in preterm and encephalopathic infants show that the bursts are characterized by long-tailed, scale-free properties that stretch across a remarkable six orders of magnitude (Roberts *et al.*, 2014a). While individual bursts are highly irregular, binning them according to their size and then averaging all bursts within each bin yields very smooth shapes that vary only slowly across scales (Roberts *et al.*, 2014a) (**Figure 5**). The relationship between these scale-specific shapes are described by a so-called *scaling function* that speaks to the nature of criticality in all systems where it arises (Baldassarri *et al.*, 2003; Papanikolaou *et al.*, 2011). Notably, the exponents that describe the scaling and shape of the bursts pre-empt clinical outcomes in these neonates (Iyer *et al.*, 2015b; Iyer *et al.*, 2014). These shapes have been modelled by a term that describes metabolic depletion and replenishment in these critically vulnerable newborn brains (Roberts *et al.*, 2014a). Despite the success in modelling BS in babies using a criticality framework, bursts in propofol anaesthesia do not have scale-free properties (Ching *et al.*, 2010). In contrast, the effect of this type of anaesthesia in brain activity has been modelled with alternative (non-critical) dynamical mechanisms (Bojak *et al.*, 2015).

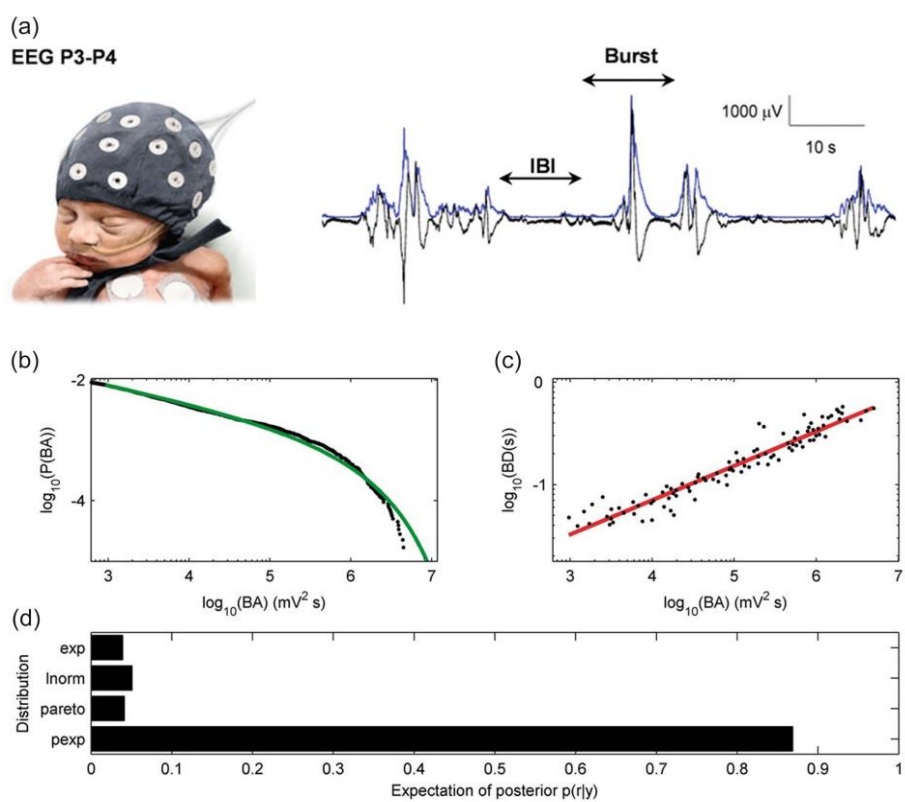

*Figure 5:* Critical statistics in neonatal hypoxic ischemic encephalopathy. (a) Scalp EEG channels show characteristic bursting pattern of activity with bursts erratically punctuating a flat EEG trace. (b) The area under each of these bursts (a measure of the energy discharged) shows power-law scaling over 3 orders of magnitude, with an exponential truncation at the far right. (c) The relationship between the length and duration of bursts also follows a power law (linear in double logarithmic coordinates. (d) Bayesian model selection shows that a truncated power law is easily the best model for these data. Panels (a) adapted from (Iyer et al., 2015a); panels (b), (c) and (d) adapted from (Roberts et al., 2014a).

Establishing the presence of critical instabilities in neurological disorders opens a number of novel therapeutic and diagnostic windows [for review, see Coombes and Terry (2012)]. For example, in concert with computational modelling, treating seizures as transitions across bifurcations permits tracking the trajectory of the seizure through bifurcation space (Freestone *et al.*, 2013; Nevado-Holgado *et al.*, 2012), with a longer-term objective of seizure control (Nelson *et al.*, 2011). Intriguingly, such control systems could use micro-stimulation to exploit the increased dynamic responsiveness of a system close to criticality as an "early warning system" (Scheffer *et al.*, 2009): To our knowledge, incorporating this approach to seizure prediction within the framework of criticality theory has not yet been achieved. Likewise, deriving measures of near-criticality from clinical, bed-side recordings of burst suppression in preterm and encephalopathic neonates could be used as prognostic markers of long-term outcome, hence opening novel therapeutic windows for early intervention (Roberts *et al.*, 2017).

6.3. Criticality and neuropsychiatric disorders

Criticality, bifurcations and phase transitions have provided increasingly nuanced understandings of several major neurological conditions. It is intriguing to consider a possible role in neuropsychiatric conditions such as psychotic and affective disorders, although these are likely to be more subtle in their deviation from the "optimum critical point". Cognitive disturbances in schizophrenia, such as working memory, have been captured in computational models by changing the influence of fluctuations in a multistable landscape (Loh *et al.*, 2007). Changes in large-scale resting-state dynamics seen in this disorder – such as a decrease in the global brain signal – have also been modelled by altering the balance of noise and stability in macroscopic neuronal activity (Yang *et al.*, 2014). Schizophrenia is associated with less suppression of activity in the default-mode network when subjects engage in external tasks (Harrison *et al.*, 2007; Meyer-Lindenberg *et al.*, 2001; Nejad *et al.*, 2011; Whitfield-Gabrieli *et al.*, 2009). Dynamic insights into this loss of efficiency could draw from the theory of criticality. In particular, if adaptive cortical activity reflects a switch from sub-critical to super-critical activity under the influence of attention and arousal, then it follows that the precision of this process could be diminished in schizophrenia. Default activity would then be inadequately suppressed during attention to external tasks, while super-critical activity would not be optimally portioned into distinct spatiotemporal patterns. While this is clearly speculative, the tools of criticality provide new methodological tools and innovative directions for research in this often-intractable disorder.

Whereas schizophrenia is characterized by disorganization – suggesting too little stability – core manifestations of other psychiatric disorders suggest too much stability. The ruminations of depressive disorders and the maladaptive pre-occupations of obsessive-compulsive disorder may arise in a system residing too deeply in the subcritical zone, preventing adaptive switching from interospective to exteroceptive states. Analysis of resting-state fMRI from patients with melancholia supports this view, showing a disturbance in the key dynamic parameters that modulate stability (Hyett *et al.*, 2015a). Notably, when patients with melancholia view emotionally salient material, frontal attentional networks increase system stability, contrary to their influence in healthy controls (Hyett *et al.*, 2015b). As noted above, critical instability is needed for a system to respond to the stream of stimuli. Here, too little instability is proposed to underlie the lack of reactivity.

Although these are preliminary findings, the potential for criticality to provide fresh insights into tough neuropsychiatric research problems is again evident. It is quite possible that psychiatric phenotypes do not only reflect a simple failure of critical brain dynamics, but also their replacement by different dynamic processes in compensation (Breakspear *et al.*, 2015).

# 7. Summary

Evidence for the widespread occurrence of criticality in nature, and its corresponding computational advantages, has triggered the interest of scientists in many different fields. The list of advantages associated with criticality spans many systems and different measurable quantities (Assis and Copelli, 2008; Boedecker *et al.*, 2012; Deco *et al.*, 2013; Gollo *et al.*, 2013; Haldeman and Beggs, 2005; Hidalgo *et al.*, 2014; Kastner *et al.*, 2015; Legenstein and Maass, 2007; Livi *et al.*, 2016; Mosqueiro and Maia, 2013; Publio *et al.*, 2012; Schrauwen *et al.*, 2009; Shew and Plenz, 2013). Despite the field traditionally developing outside of neuroscience, many of the most exciting findings now focus on brain dynamics. This body of work suggests that specific functions of the central nervous system may exploit optimal properties observed at criticality, representing a general functional property of the brain (Chialvo, 2006).

What are the mechanisms that establish criticality in the brain? The obvious lack of an externally tunable parameter highlights the importance of self-organized criticality (SOC). As we reviewed above, SOC arises when the interactions between the elements (neurons) of a system are endowed with plasticity, or when there is a slow build-up and a fast release of energy. Synaptic processes such as frequency adaptation or spike-time dependent plasticity are clear candidates for SOC at the neuronal level. Homeostatic mechanisms, such as the hypothalamus-pituitary axis, govern metabolic processes in the body, and reflect the slow time-scales of energy accumulation and utilization. In this vein, neuronal criticality may extend beyond the central nervous system to incorporate the autonomic nervous system and corresponding interoceptive feedback loops to the midbrain and insula. Although criticality can arise from a fairly simple memory process, it seems unlikely to be confined to any single mechanism in the brain, but may reflect multiple processes operating for different goals (computation, adaptation, homeostasis). Likewise, changes in the environment, such as a scarcity of food sources, which mandate an (allostatic) change in the balance of energy metabolism may also engage SOC processes through a change in the critical set point. These more complex feedback systems are not classically thought of as SOC, but rather homeostatic, self-regulatory systems of the brain and body. Future work is required to understand if these broader homeo- and allostatic processes do follow the principles of SOC.

Conversely, pathological failures of adaptive criticality may reflect disturbances in any of the supporting mechanisms, such as when a failure of balanced inhibition leads to a seizure, or when interoceptive disturbances lead to fatigue and depression (Petzschner *et al.*, 2017). Criticality may also arise in pathological settings, such as if metabolic disturbances lead to an abnormal build-up and release of energy that would not normally occur: Burst suppression in the hypoxic newborn may be an example of this scenario (Roberts *et al.*, 2015).

The study of the brain as a complex, dynamic network is now flourishing. Modelling the inherently unstable, multiscale neuronal dynamics that generate flexible cognitive states is crucial to this endeavour. Although the notion of criticality is an attractive candidate, much work remains to be done to clarify the relationship between cognition and criticality. The increasing sophistication of theoretical models and their hypothesis-driven application to empirical data has to be tempered against the caveats that have recently emerged. Research that incorporates these lessons will shed new light on the dynamic brain networks that underlie action, perception and cognition in health and disease.

Unifying models of brain function give sense to the results generated by the boom of explorative studies assessing the association between cognitive functions and neural dynamics. Without such models, there is a risk that the accumulation of experimental evidence will not substantially impact upon our understanding of the neural underpinnings of human cognition in health and disease. Recent studies suggest that the transition from critical to super-critical regimes may represent a general principle underlying the emergence of goal-directed behaviour. Those transitions may be moderated by slow integrators that are the highly connected hubs of the brain. In general, the framework of criticality challenges earlier prevailing conceptualizations of normal and pathological cognitive functions as emerging from discrete regions or static networks of brain regions.

The science of criticality provides a new armoury of analytic techniques for basic and translational neuroscientists. The field is now sufficiently mature that inferring the presence of scale-free statistics from time series data should derive from formal likelihood tests and comparisons to other heavy-tailed processes. In addition, showing that the spatial and/or temporal statistics of a system are scale-free is now only the first step in emerging "best practice". Inferring that the underlying system is critical should ideally rest upon showing that the best model that can account for those statistics is one in which the model itself is perched at the cusp of criticality. Criticality should not become a catch-all term for everything that is complex or variable. It is only one amongst many possible causes of complex dynamics and care should be taken to disambiguate amongst these.

Funding: L.C., L.L.G, A.Z and M.B. were supported by the Australian National Health Medical Research Council (L.C. APP1099082, L.L.G. APP1110975, A.Z. APP1047648, M.B. APP1037196). This work was also supported by the Australian Research Council Centre of Excellence for Integrative Brain Function (M.B., ARC Centre Grant CE140100007).

Conflicts of Interest: The authors declare that the research was conducted in the absence of any commercial or financial relationships that could be construed as a potential conflict of interest.

Acknowledgments: Dedicated to the memory of Walter Freeman whose seminal early work inspired us and shaped the field.

# Box 1: Ten-Point Summary

1. Criticality arises when a system is close to dynamic instability and is reflected by scale-free temporal and spatial fluctuations

2. Critical temporal fluctuations (crackling noise) occur in simple systems close to a bifurcation

3. Critical spatiotemporal fluctuations (avalanches) occur in complex systems close to a phase transition

4. Crackling noise and avalanches have now been observed in a wide variety of neuronal recordings, at different scales, in different species, and in health and disease

5. Computational models suggest a host of adaptive benefits of criticality, including maximum dynamic range, optimal information capacity, storage and transmission and selective enhancement of weak inputs

6. Resting-state EEG and fMRI data show evidence of critical dynamics

7. The onset of a specific cognitive function may reflect the stabilization of a particular subcritical state under the influence of sustained attention

8. Mounting evidence and models suggest that several neurological disorders such as epilepsies and neonatal encephalopathy reflect bifurcations and phase transitions to pathological states

9. Novel insights into neuropsychiatric disorders such as schizophrenia and melancholia might also be obtained by leveraging the tools of criticality, although this currently remains somewhat speculative

10. While the application of criticality to neuroscience is an exciting field, progress needs to proceed with due caution, using appropriate methods, considering alternative complex processes and using computational models in partnership with data analysis

# Box 2: Glossary

The *attractor* of a dynamical system is the set of all points traversed once initial transients have passed. Attractors can be *fixed points* (with steady state solutions), *limit cycles* (periodic) or *chaotic* (deterministic but dynamically unstable and aperiodic).

The *basin of attraction* of an attractor is the set of all initial conditions that eventually flow onto that attractor.

A system can have more than one attractor even if all its parameters are fixed. Such a system is said to be *multistable*. Such systems will also have multiple basins of attraction, separated by *basin boundaries. Bistability occurs* in a multistable system which has exactly two attractors.

A *metastable* system does not have any attractors. It instead has a series of saddles (fixed points with attracting and repelling subspaces) that are linked into a complex (heteroclinic) cycle. A metastable system will jump endlessly from the neighborhood of one saddle to another. Metastable systems are also called *winnerless competition.*

An attractor that only changes slightly (and smoothly) when its underlying parameters are changes is said to be *structurally stable*. If the topology (shape) of the attractor fundamentally changes then the attractor is said to be *unstable* – or, alternatively undergo a *bifurcation*. The value of the parameter at which that discontinuous change occurs is said to be a *critical* or *bifurcation point.*

A system consisting of many interacting components may exhibit a sudden change in state in the presence of a slowly tuned *control parameter* (such as temperature). Such a transition is called a *phase transition* and typically separates an ordered from a disordered state. Technically, a phase transition corresponds to a discontinuity in the thermodynamic free energy of a system.

*Criticality* occurs when a system is poised at the point of a dynamic instability. Because of this, microscopic fluctuations are not damped but instead appear at all scales of the system. This yields power-law fluctuations in the temporal domain ("crackling noise") and the spatiotemporal domain ("avalanches"). A critical system will show scaling laws, such that a single (universal) function can map the shape of fluctuations at any scale into those at the scale above (or below).

Bifurcations may be *super-critical* (when stable oscillations appear above the critical point) or *sub-critical* (when a zone of bistability occurs below the critical point). Only supercritical bifurcations can yield power law (critical) fluctuations: Sub-critical bifurcations lead to multistable switches that occur on a characteristic time scale.

Phase transitions may be continuous (second order) or discontinuous (first order). Mathematically, these are equivalent to a super- and sub-critical bifurcations, respectively. As with bifurcations, critical power law fluctuations only occur in the neighbourhood of a continuous phase transition. Discontinuous phase transitions can yield multistable switching or complex mixtures of states (such as in boiling water).

Some systems need to be externally tuned by a control parameter close to their critical point. In other systems, criticality will emerge from many initial parameter values, usually due to plasticity and memory.

The *order parameter* of a complex system is a macroscopic observable such as the magnetic field of a ferromagnet. A non-zero order parameter arises in the ordered state of a system, in the supercritical (or active) phase. In the subcritical phase, the order parameter remains at zero even with the addition of energy. Such a state is called an *absorbing state.*

A *power law* exists between two variables *x* and *y* if they obey the relationship: $y \propto x^{-k}$. If the power law arises in the setting of criticality, then the constant *k* is called the critical exponent. A system shows power law behaviour if the probability density function of its fluctuations obeys $f(x) \propto x^{-k}$ for all values of *x* greater than some minimum cut-off $x_{min}$. A power law probability distribution is also called a *Pareto* distribution. The corresponding cumulative distribution obeys $Prob(X > x) \propto 1 - (x/x_{min})^{-k+1}$ for the same critical exponent *k*.

A system is said to be *scale-free* when it doesn't have a characteristic time or length scale. Scale-free systems show a power law probability distribution over several orders of magnitude with an exponent *k* that is less than 2. Correspondingly, the variance of a scale-free system is only bounded by the system size.

A complex system shows evidence of *slowing down* when it is close to a critical point – that is, the time scale of its fluctuations (the characteristic return to the mean) slow down, changing from a fast (exponential) process to a slow power law.

A system has an *exponential distribution* when its probability density function is given by $f(x) \propto e^{-x/L}$. The system has a single time scale, corresponding to the constant exponent *L*.

Any distribution whose probability distribution drops off more slowly than an exponential distribution is said to be heavy-tailed. The Pareto distribution $f(x) \propto x^{-k}$ is a classic heavy tailed distribution but log-normal and stretched exponential (Weibull) distributions are also heavy-tailed.

Many physical systems close to a critical point do have some weak damping that acts on the largest fluctuations: These systems show an *exponentially-truncated power law*, $f(x) \propto x^{-k} e^{-xl}$.